\documentclass[twocolumn,showpacs,preprintnumbers,amsmath,amssymb,nofootinbib,superscriptaddress]{revtex4-1}
\usepackage{graphicx}
\usepackage{dcolumn}
\usepackage{bm}

\usepackage{hyperref}
\usepackage{tikz}
\usepackage{tikz-3dplot}
\usetikzlibrary{decorations.markings}

\hypersetup{
    pdftitle={On Discrete Symmetries and Torsion Homology in F-Theory},
    pdfauthor={Christoph Mayrhofer, Eran Palti, Oskar Till, Timo Weigand},
    pdfsubject={F-theory/M-theory compactifications}
}

\newcommand{\be}{\begin{equation}}
\newcommand{\ee}{\end{equation}}
\newcommand{\bea}{\begin{eqnarray}}
\newcommand{\eea}{\end{eqnarray}}

\newcommand{\nn}{\nonumber}

\newcommand{\tw}{{\rm w}}

\begin{document}
\title{On Discrete Symmetries and Torsion Homology in F-Theory}
\author{Christoph Mayrhofer}
\affiliation{Arnold-Sommerfeld-Center, Ludwig-Maximilians-Universit\"at M\"unchen, Germany}
\author{Eran Palti}
\affiliation{Institut f\"ur Theoretische Physik, Ruprecht-Karls-Universit\"at Heidelberg, Germany}
\author{Oskar Till}
\affiliation{Institut f\"ur Theoretische Physik, Ruprecht-Karls-Universit\"at Heidelberg, Germany}
\author{Timo Weigand}
\affiliation{Institut f\"ur Theoretische Physik, Ruprecht-Karls-Universit\"at Heidelberg, Germany}
\begin{abstract}
We study the relation between discrete gauge symmetries in F-theory compactifications and torsion homology on the associated Calabi-Yau manifold. Focusing on the simplest example of a $\mathbb Z_2$ symmetry, we show that there are two physically distinct ways that such a discrete gauge symmetry can arise. First, compactifications of M-Theory on Calabi-Yau threefolds which support a genus-one fibration with a bi-section are known to be dual to six-dimensional F-theory vacua with a $\mathbb Z_2$ gauge symmetry. We show that the resulting five-dimensional theories do not have a $\mathbb Z_2$ symmetry but that the latter emerges only in the F-theory decompactification limit. Accordingly the genus-one fibred Calabi-Yau manifolds do not exhibit torsion in homology. Associated to the bi-section fibration is a Jacobian fibration which does support a section. Compactifying on these related but distinct varieties does lead to a $\mathbb Z_2$ symmetry in five dimensions and, accordingly, we find explicitly an associated torsion cycle. We identify the expected particle and membrane system of the discrete symmetry in terms of wrapped M2 and M5 branes and present a field-theory description of the physics for both cases in terms of circle reductions of six-dimensional theories. Our results and methods generalise straightforwardly to larger discrete symmetries and to four-dimensional compactifications.
\end{abstract}
\maketitle


\section{Introduction}

Discrete symmetries not only play a prominent role in particle physics as selection rules governing the structure of interactions, but they are also interesting  by themselves from the perspective of quantum field theory.
Indeed, the conjecture that in quantum gravity no global continuous symmetries exist is believed to apply also to global discrete symmetries \cite{Banks:2010zn}.
String theory as a candidate for a quantum theory of gravity is therefore a natural arena to study discrete symmetries and their emergence from gauge symmetries at high energy scales as believed to be required for consistency with black hole physics.
This question has received a lot of recent attention in the context of compactifications of type II string and M-theory \cite{Camara:2011jg,BerasaluceGonzalez:2011wy,Ibanez:2012wg,BerasaluceGonzalez:2012vb,Anastasopoulos:2012zu,BerasaluceGonzalez:2012zn,Honecker:2013hda,Berasaluce-Gonzalez:2013sna,Berasaluce-Gonzalez:2013bba}.
If we focus for definiteness on type IIA compactifications on a Calabi-Yau 3-fold $X_3$, the appearance of a closed string Ramond-Ramond (RR) $\mathbb Z_k$ symmetry is in one-to-one correspondence with the existence of
torsional (co)homology groups on $X_3$ \cite{Camara:2011jg}.
Indeed, the smoking gun for a $\mathbb Z_k$ symmetry in four-dimensional field theory is the existence of $\mathbb Z_k$ charged particles and strings \cite{Banks:2010zn}. While in field theory these arise a priori as 
operators describing the associated probe particles and strings, in quantum gravity all such operators are conjectured to be realized a forteriori as physical objects. 
In type IIA compactifications these $\mathbb Z_k$ charged particles and strings are due to wrapped D2- and D4-branes along $k$-torsional 2- and 3-cycles. By definition, $k$ copies of such $k$-torsional cycles are homologically trivial, in agreement with the fact that $k$ copies of the $\mathbb Z_k$ charged particles and strings are uncharged and can thus decay \cite{Camara:2011jg}. 
Furthermore, the existence of such torsional 2- and 3-cycles implies the existence of a torsional 3-form $\alpha$ with the property
\begin{equation}
k \, \alpha = d \tw.
\end{equation}
Dimensional reduction of the RR 3-form $C_3$ as $C_3 = A \wedge \tw + \ldots $ gives rise to a massive $U(1)$ gauge potential $A$ whose associated gauge symmetry is in fact broken to $\mathbb Z_k$, which is precisely the discrete symmetry observed in the effective action. 
Thus a closed string $\mathbb Z_k$ symmetry in type IIA on $X_3$ manifests itself geometrically in the fact that \cite{Camara:2011jg}
\begin{equation}
\begin{aligned}
& {\rm Tor} H_2(X_3, \mathbb Z) \simeq {\rm Tor} H_3(X_3, \mathbb Z)  = \mathbb Z_k, \\
& {\rm Tor} H^3(X_3,\mathbb Z) \simeq  {\rm Tor} H^4(X_3,\mathbb Z) =  \mathbb Z_k.
\end{aligned}
\end{equation}

More recently, the origin of discrete symmetries has been studied in the framework of F-theory compactifications  \cite{Braun:2014oya,Morrison:2014era,Anderson:2014yva,Klevers:2014bqa,Garcia-Etxebarria:2014qua,Mayrhofer:2014haa}.\footnote{Non-abelian discrete symmetries in F-theory spectral cover models have been discussed in \cite{Karozas:2014aha} and references therein.} One way how discrete symmetries arise turns out to be from F-theory compactifications on genus-one fibrations without a section \cite{Braun:2014oya}. More precisely, if a genus-one fibration possesses only a multi-section of order $k$ (as opposed to a rational section), this manifests itself generically in a $\mathbb Z_k$ symmetry of the F-theory effective action.
In view of the geometric realization of closed string discrete symmetries in type II theory sketched above, a natural question is whether torsional (co)homology plays any role in this picture.

In this note we answer this question and along the way
clarify a number of open questions and puzzles concerning both the geometry and the field theory underlying the class of discrete symmetries in F-theory studied so far in \cite{Braun:2014oya,Morrison:2014era,Anderson:2014yva,Klevers:2014bqa,Garcia-Etxebarria:2014qua,Mayrhofer:2014haa}. 
For concreteness we will be working in the context of the simplest possible type of discrete symmetry, a $\mathbb Z_2$ symmetry, but our conclusions immediately extend to larger discrete symmetry groups. 

Associated with a $\mathbb Z_2$ symmetry in the effective action of F-theory compactified to $2n$ large dimensions is a pair of related fibrations $P_Q$ and $P_W$ which are genus-one fibrations over a base $B_{5-n}$ of complex dimension $5-n$. 
The fibration $P_Q$ only possesses a bi-section \cite{Braun:2014oya}, while $P_W$ is the singular Jacobian of $P_Q$ which does exhibit a zero-section. 
More generally, a $\mathbb Z_k$ symmetry is related to a set of $k$ isomorphism classes of genus-one fibrations with the same Jacobian, classified by the Tate-Shafarevich group of the latter.
As we will discuss in great detail extending the results of \cite{Morrison:2014era,Mayrhofer:2014haa}, both fibrations give rise to identical compactifications in $2n$ dimensions, but when compactifying M-theory on either of them to $2n-1$ dimensions, the theories are strikingly different. This difference of the underlying M-theory compactification will be shown to correspond, amongst other things, to the presence of torsional (co)homology on the Jacobian fibration $P_W$, which is absent on $P_Q$.  
For simplicity we will mostly focus on the case $n=3$ in the sequel (with the exception of the end of section \ref{sec:PQ}), which corresponds to F-theory and M-theory  compactifcations to six and five dimensions, respectively. We will see how the two different M-theory compactifications give rise to the same effective F-theory model in six dimensions by taking into account that the Higgs field which breaks to the $\mathbb Z_2$ symmetry can have a spatially varying vacuum expectation value along the circle relating the six and five-dimensional theories. This will offer a dual perspective on, and allow us to derive directly quantitative properties of, the observation first made in \cite{Witten:1996bn}, and subsequently substantially developed further in \cite{Anderson:2014yva,Garcia-Etxebarria:2014qua}, that F-theory compactifications without a section are dual in terms of a fluxed circle reduction to M-theory. 

This field theoretic picture for F/M-theory on a genus-one fibration $P_Q$ without section will be developed in chapter \ref{sec:PQ}. At the end of this section we also comment further on the $G_4$ flux in four-dimensional compactifications introduced for these models in \cite{Mayrhofer:2014haa}. We then analyse in section \ref{sec:PW} the different geometric and physical properties of F/M-theory on the Jacobian fibration $P_W$. We explictly identify torsional 3-cycles and analyse a birational blow-up of $P_W$ which allows us to deduce also the expected torsional 2-cycles. More details on this computation can be found in the \hyperlink{sec:blow-up}{appendix}.

\section{$\mathbb Z_2$ symmetry from genus-one fibrations} \label{sec:PQ}

The first type of geometry, $P_Q$, is a genus-one fibration whose fibre takes the form of a quartic hypersurface in $\mathbb P_{112}$ with homogenous coordinates $[u : v : w]$,
\begin{equation}
\begin{aligned} \label{PQ}
P_Q =&  w^2 + b_0 u^2 w  + b_1 u v w + b_2 v^2 w  + c_0 u^4   \\ 
& + c_1 u^3 v  +  c_2 u^2 v^2  + c_3 u v^3  + c_4 v^4.
\end{aligned}
\end{equation}
The coefficients are sections of suitable line bundles over the base $B_2$, which we take, for the time-being, to be two-complex-dimensional.
As first discussed in detail in \cite{Braun:2014oya}, this fibration does not possess a section, but only a bi-section 
\begin{equation}
U_{\rm bi}: \{u= 0 \}
\end{equation}
 intersecting the generic fibre in two points exchanged by monodromies along a branch  cut on $B_2$. 
For generic coefficients the fibration contains a smooth $I_2$-fibre over a specific co-dimension-two locus $C$ on the base $B_2$.
Each of the two fibre components $A_C$ and $B_C$ over $C$ are intersected by the bi-section once.
M2-branes wrapping the two fibre components  give rise, in the F-theory limit, to massless states with $\mathbb Z_2$-charge $1$ mod $2$. 
This $\mathbb Z_2$ charge manifests itself explicitly as a selection rule governing the Lagrangian and becomes particularly effective in the presence of extra non-abelian gauge groups \cite{Morrison:2014era,Anderson:2014yva,Klevers:2014bqa,Garcia-Etxebarria:2014qua,Mayrhofer:2014haa}.

\begin{figure}
\centering
\begin{tikzpicture}
\node at (1.3,2) {fibre over $C_I$};
\node at (-0.2,1) {$A_I$};
\node at (2.8,1) {$B_I$};
\begin{scope}[scale=0.7]
\begin{scope}[xshift=3.6cm]
   \shadedraw[ball color=gray!30!white,xscale=-1] (-0.3,0) .. controls (-0.3,1) and (.3,2) .. (1,2)
               .. controls (1.7,2) and (1.8,1.5) .. (1.8,1.2)
               .. controls (1.8,.5) and (1,.5) .. (1,0)
               .. controls (1,-.5) and (1.8,-.5) .. (1.8,-1.2)
               .. controls (1.8,-1.5) and (1.7,-2) .. (1,-2)
               .. controls (.3,-2) and (-0.3,-1) .. (-0.3,0);
\end{scope}
\shadedraw[ball color=blue!40!white] (-0.3,0) .. controls (-0.3,1) and (.3,2) .. (1,2)
               .. controls (1.7,2) and (1.8,1.5) .. (1.8,1.2)
               .. controls (1.8,.5) and (1,.5) .. (1,0)
               .. controls (1,-.5) and (1.8,-.5) .. (1.8,-1.2)
               .. controls (1.8,-1.5) and (1.7,-2) .. (1,-2)
               .. controls (.3,-2) and (-0.3,-1) .. (-0.3,0);
\draw[green, line width=0.5mm, xshift=.5cm,yshift=1cm] (0,0)--(0.2,0.2);
\draw[green, line width=0.5mm, xshift=.5cm,yshift=1cm] (0,0.2)--(0.2,0);
\end{scope}

\begin{scope}[xshift=4cm]
\node at (1.3,2) {fibre over $C_{II}$};
\node at (-0.3,1) {$A_{II}$};
\node at (2.8,1) {$B_{II}$};
\begin{scope}[scale=0.7]
   \shadedraw[ball color=gray!30!white] (-0.3,0) .. controls (-0.3,1) and (.3,2) .. (1,2)
               .. controls (1.7,2) and (1.8,1.5) .. (1.8,1.2)
               .. controls (1.8,.5) and (1,.5) .. (1,0)
               .. controls (1,-.5) and (1.8,-.5) .. (1.8,-1.2)
               .. controls (1.8,-1.5) and (1.7,-2) .. (1,-2)
               .. controls (.3,-2) and (-0.3,-1) .. (-0.3,0);
\begin{scope}[xshift=3.6cm]
\shadedraw[ball color=gray!80!white,xscale=-1] (-0.3,0) .. controls (-0.3,1) and (.3,2) .. (1,2)
               .. controls (1.7,2) and (1.8,1.5) .. (1.8,1.2)
               .. controls (1.8,.5) and (1,.5) .. (1,0)
               .. controls (1,-.5) and (1.8,-.5) .. (1.8,-1.2)
               .. controls (1.8,-1.5) and (1.7,-2) .. (1,-2)
               .. controls (.3,-2) and (-0.3,-1) .. (-0.3,0);
\end{scope}
\draw[green, line width=0.5mm, xshift=1cm,yshift=1cm] (0,0)--(0.2,0.2);
\draw[green, line width=0.5mm, xshift=1cm,yshift=1cm] (0,0.2)--(0.2,0);
\draw[blue!40!white, line width=0.5mm, xshift=2.5cm,yshift=-1cm] (0,0)--(0.2,0.2);
\draw[blue!40!white, line width=0.5mm, xshift=2.5cm,yshift=-1cm] (0,0.2)--(0.2,0);
\end{scope}
\end{scope}

\end{tikzpicture}
\caption{The fibre structure over the singlet curves $C_I$ and $C_{II}$ taken from \cite{Mayrhofer:2014haa} with blue denoting  the section $S$ and green  the section $U$.}\label{fig:fibre}
 \end{figure}
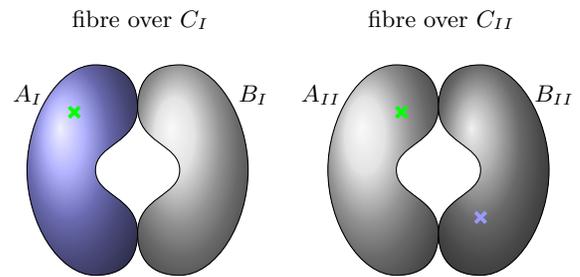

The appearance of a $\mathbb Z_2$-symmetry in the six-dimensional effective theory has been understood in \cite{Morrison:2014era,Anderson:2014yva,Klevers:2014bqa,Garcia-Etxebarria:2014qua,Mayrhofer:2014haa} as the effect of the Higgsing of a six-dimensional $U(1)$ gauge symmetry with a Higgs field of charge $2$.\footnote{See \cite{Grassi:2014sda} for a recent discussion of non-abelian Higgsing in F-theory.} Let us briefly review this six-dimensional picture. The unhiggsed theory arises by F-theory compactification on a related elliptic fibration $\hat P_Q$ with Mordell-Weil group of rank one, given by \cite{Morrison:2012ei}
\begin{equation}
\begin{aligned} \label{eq:Bl1124}
\hat P_Q =& s w^2   + b_{0}  s^2 u^2 w  + b_1s u v w  + b_2 v^2 w + \\
&  + c_{0}s^3 u^4   + c_{1} s^2  u^3 v  + c_{2}s u^2 v^2   + c_3 u v^3.
\end{aligned}
\end{equation}
Compared to (\ref{PQ}), the coefficient $c_4$ vanishes; the hypersurface ${P_Q}|_{c_4 \equiv 0}$ acquires a conifold singularity along the co-dimension four locus $w=u=b_2 = c_3 =0$. These conifold singularities admit a small resolution by blowing up the ambient space, thereby introducing the exceptional divisor $S: s=0$. The resulting fibration $\hat P_Q$ is therefore described by a ${\rm Bl}^1 \mathbb P_{112}[4]$-fibration over the base and has two independent rational sections $S$ and $U$. To avoid confusion we will reserve the notation $U$ for the section $U: u=0$ on $\hat P_Q$, in contrast to the bi-section $U_{\rm bi}$ on $P_Q$. 
In F-theory on $\hat P_Q$, the six-dimensional $U(1)$ symmetry arises by duality with M-theory upon expanding $C_3 = A \wedge \tw$ with $\tw$ the Poincar\'e dual of the image of the rational section $S$ under the Shioda map, where the image of $S$ is $S-U - (b_2 + \bar {\cal K})$.
The fibre splits over two different co-dimension-two loci $C_I$ and $C_{II}$ into two rational curves  ${A_I}$, ${B_I}$ and, respectively,  ${A_{II}}$, ${B_{II}}$ as depicted in FIG.~\ref{fig:fibre}.
Our  notation is such that the fibre components $A_I$ and $A_{II}$ are intersected by the section $U$. The locus $C_I = \{b_2 = 0\} \cap \{c_3 = 0\}$ is the locus of the conifold singularities in ${P_Q}|_{c_4 \equiv 0}$, while the locus $C_{II}$ is associated with a more complicated prime ideal. Note that the locus $C$ on $P_Q$ is the remnant of $C_{II}$ after switching on $c_4$. The intersection numbers of the fibre components with the independent sections $S$ and $U$ are also summarized in TABLE~\ref{hatPQcharges}.
\begin{table}
\begin{tabular}{c|| c|c || c | c }
               &  $A_I$ & $B_I$ & $A_{II}$ & $B_{II}$ \\ \hline \hline
$S$        &  -1     &  2  &      0     &      1         \\
$U$        &  1     &    0    &      1     &       0          \\
$S+U$    &  0     &   2   &      1      &     1               \\
$S-U$     &  -2     &   2  &      -1     &     1
\end{tabular}
\caption{$U(1)$ charges of M2-branes wrapping fibre components in M-theory compactified on $\hat P_Q$. \label{hatPQcharges}}
\end{table}
The intersection number with $S-U$ is the charge of M2-branes wrapping the various fibre components with respect to the six-dimensional massless $U(1)$ gauge symmetry. Thus, 
the Higgs field of charge $\pm 2$ with respect to the six-dimensional $U(1)$ symmetry is associated with an M2-brane wrapping either of the fibre components over the locus $C_I$. The conifold transition from $\hat P_Q$ to $P_Q$ corresponds to giving the Higgs field with charge $-2$ a VEV, thereby  breaking the six-dimensional $U(1)$ gauge symmetry to $\mathbb Z_2$.

The appearance of this $\mathbb Z_2$ discrete symmetry in F-theory as reviewed above leads to a puzzle: Following the field theoretic arguments of \cite{Banks:2010zn}, we expect sets of $\mathbb Z_2$ charged particles and, since we are working in six spacetime dimensions, also sets of dual (3+1)-dimensional operators. Adapting the logic of \cite{Camara:2011jg} to the case at hand these should come from M2- and M5-branes wrapped on torsional 2- and 3-cycles, respectively. 
However, for generic base spaces $B_2$ the fibration $P_Q$ does not possess any such torsion elements as follows from the general analysis of integral cohomologies of toric hypersurfaces in \cite{BatyrevKreuzer}.

We now resolve the puzzle about the absence of torsion on $P_Q$ by carefully identifying the various $U(1)$ symmetries and the Higgs field, in particular regarding the five-dimensional versus six-dimensional fields. Let us consider M-theory compactified on $\hat P_Q$ to five spacetime dimensions. 
In five dimensions, both independent sections $S$ and $U$ give rise to a bona fide $U(1)$ gauge symmetry from expanding $C_3$.\footnote{We ignore further $U(1)$ symmetries due to elements of $H^{1,1}(B_2)$ as these play no role in our considerations.} The intersection numbers in TABLE~\ref{hatPQcharges} then compute the charges of M2-branes wrapping the various fibre components over $C_I$ and $C_{II}$.

The deformation process that relates $\hat P_Q$ with $P_Q$ corresponds to first blowing down the exceptional divisor $S$ and then switching on $c_4$,
\begin{equation}
\hat P_Q \rightarrow P_Q|_{c_4 \equiv 0} \rightarrow P_Q. \label{bdq}
\end{equation}
Importantly, on the singular fibration $P_Q|_{c_4 \equiv 0}$ the fibre $A_I$ shrinks to zero size while all other fibres remain of finite volume. In particular the volume of $A_{II}$ and $B_{II}$ are equal and non-zero \cite{Morrison:2014era,Mayrhofer:2014haa} and, as further stressed in \cite{Mayrhofer:2014haa}, it is the M2-branes wrapping $A_I$ that become massless. The associated hypermultiplet serves as the massless Higgs field which obtains a VEV upon switching on $c_4$. Since the six-dimensional $U(1)$ corresponds to the combination $S-U$ and the Higgs field is uncharged under $S+U$, one might be tempted to work in the basis $U(1)_{S-U}$ and $U(1)_{S+U}$ and  naively conclude that 
in the five-dimensional field theory this deformation higgses $U(1)_{S+U} \times U(1)_{S-U} \rightarrow U(1)_{S+U} \times \mathbb Z_2$. However the correct prescription for determining the discrete symmetry is to
bring the Higgs, via a unimodular transformation, into a basis where it is only charged under a single $U(1)$, cf.~appendix~A of \cite{Mayrhofer:2014haa}.
It is simple to check that 
the appropriate 
basis is $U(1)_{S+U}$ and $U(1)_U$. Since the Higgs field has charge $(0,1)$ under these symmetries, the gauge symmetry breaking  in five dimensions is
\bea\label{U1breakingPQ}
U(1)_U \times U(1)_{S+U} \rightarrow U(1)_{S+U}.
\eea
This explains the absence of $\mathbb Z_2$-torsion (co)homology on $P_Q$: In the five-dimensional effective theory obtained by dimensional reduction of M-theory on $P_Q$ no discrete gauge symmetry arises. Similarly, and again consistently with absence of torsion on $P_Q$, compactification of type II theory on $P_Q$ gives rise to a four-dimensional ${\cal N}=2$ effective theory with gauge group $U(1)_{S+U}$ and no discrete RR symmetry.

This conclusion, in turn, calls for an explanation of the fact that in the six-dimensional F-theory compactification on $P_Q$ there {\emph{is}} a $\mathbb Z_2$ symmetry while in the five-dimensional M-theory compactification the gauge group is $U(1)_{S+U}$. To understand this one must analyse the 
 reduction of the six-dimensional theory on a circle. Consider the five-dimensional theory obtained from M-theory on $\hat P_Q$, i.e.\ before the Higgsing. This theory has two $U(1)$ symmetries $U(1)_U$ and $U(1)_{S-U}$. This particular basis is the appropriate one to interpret $U(1)_U$ as the Kaluza-Klein $U(1)$ coming from the metric upon reducing the six-dimensional theory on a circle and $U(1)_{S-U}$ as the zero mode of the six-dimensional $U(1)$: Indeed the tower of 
states constructed by wrapping the full elliptic fibre $n$ times generates the full integer charges for $U(1)_U$ and all have equal charges under $U(1)_{S-U}$, as expected from a KK tower. Now since the Higgs field has charge 1 under $U(1)_U$ it is a first excited KK mode (see \cite{Anderson:2014yva,Garcia-Etxebarria:2014qua,Mayrhofer:2014haa} for discussions on this point). The important point is that this implies that the background has a vacuum expectation value for the Higgs which is spatially varying along the circle. This mixes the geometric action on the wavefunctions associated to translations along the circle with the internal gauge symmetry. Since it is a first excited KK mode but has charge $-2$ under the six-dimensional $U(1)_{S-U}$ the remaining symmetry 
\bea
U(1)_{S+U} = U(1)_{S-U} + 2 U(1)_{U}
\eea 
corresponds to moving at twice the rate along the circle as along the internal $U(1)$. In particular, it means that the $\mathbb Z_2$ subgroup of $U(1)_{S-U}$, i.e.\ a shift in phase by $\pi$, takes us a full path around the circle. The $\mathbb Z_2$ is therefore actually a five-dimensional symmetry and becomes a subgroup of the remnant five-dimensional symmetry $U(1)_{S+U}$ constructed from $U(1)_U$ and the zero mode of $U(1)_{S-U}$. 
To understand why in the six-dimensional theory only the $\mathbb Z_2$ symmetry remains, consider
decompactification of the circle to a line. We should think of the action of going around the circle part of the $\mathbb Z_2$ as a map to the point at infinity added to the real line to make the circle. Upon decompactification this point is removed and what remains is just the action of the $\mathbb Z_2$ subgroup of the six-dimensional $U(1)_{S-U}$. Another way to think about it is that the wavefunction of the first KK mode becomes flat and we have the usual Higgsing of a $U(1)$ with a constant vacuum expectation value.

A related way to see the emergence of a $\mathbb Z_2$ symmetry in F-theory is as follows: A $\mathbb Z_2$ symmetry means that 
two copies of a state of charge 1 can decay to the vacuum. Consider two copies of the state associated with an M2-brane along $A_{C}$ on $P_Q$, where we recall that on $P_Q$, over the point set $C$ the fiber splits into two fiber components $A_C$ and $B_C$. These are related to $A_{II}$ and $B_{II}$ on $\hat P_Q$.
 In homology $[A_{C}] = [B_{C}]$ and thus this pair of states is equivalent to an M2-brane along $[A_C]+[B_C] = [T^2]$.
From the perspective of five-dimensional M-theory a state along $T^2$ carries KK charge  and is thus different from the vacuum, while in the six-dimensional F-theory such a state is equivalent to the vacuum. Put differently, the relation  $[A_{C}] = [B_{C}]$ implies that $2[A_C] = [T^2]$, i.e.\ $A_{C}$ is 2-torsion in $H_2(P_Q,\mathbb Z)/[T^2]$.
Thus while no torsion arises in $H_2(P_Q,\mathbb Z)$ on a bi-section fibration such as $P_Q$, torsion \emph{modulo the fiber class} does appear and guarantees a $\mathbb Z_2$ symmetry in F-theory, but not in M-theory.\footnote{This is to be contrasted with the effect of $k$-torsional elements in the Mordell-Weil group of rational sections, which, as shown in \cite{Mayrhofer:2014opa}, give rise to $k$-torsional \emph{divisors} modulo certain resolution divisors associated with the appearance non-abelian gauge symmetry.}

The above picture of Higgsing has a nice reformulation in terms of a St\"uckelberg mechanism. The usual way to write the Higgs field is as a modulus and a phase, $\phi = h e^{ic}$, where the phase part is associated with an axion $c$. Since the Higgs is a first KK mode, it depends on the circle coordinate $y$ as $e^{iy}$, which implies a linear profile for the axion field. The field therefore has an associated flux when integrated over the circle. The fact that the F-theory T-dual perspective to the M-theory geometry should be a fluxed reduction along a circle was first noted in \cite{Witten:1996bn}. More recently, in \cite{Anderson:2014yva,Garcia-Etxebarria:2014qua}, this was developed further and applied to bi-section models. In particular the duality with the flux reduction was shown to reproduce the expected lower-dimensional physics including the Chern-Simons terms. The flux then breaks the KK $U(1)_U$ while the fact that the Higgs has charge $-2$ under the six-dimensional $U(1)_{S-U}$ means that the axion couples to it with coefficient 2 and (linearly) breaks it. The resulting five-dimensional $U(1)$ is then the combination that remains of the zero mode of $U(1)_{S-U}$ and $U(1)_U$ as discussed above.

In the next section we will present the M-theory geometric perspective on this breaking and the particular point of why in M-theory no $\mathbb Z_2$ symmetry remains, while in F-theory a $\mathbb Z_2$ symmetry does remain. To summarise there are 3 dual perspectives on the same physics of the breaking: the linear Higgs field theory (open string picture), the non-linear or St\"uckelberg mechanism (closed string picture)\footnote{In \cite{Anderson:2014yva,Garcia-Etxebarria:2014qua} a map from the axion to the closed-string sector of IIB has been proposed.}, and the M-theory geometry (T-dual picture). Note that these are really dual rather than co-existing effects.

Our analysis immediately generalises to compactifications of F/M-theory to four and, respectively, three dimensions on fibrations over a three-complex dimensional base $B_3$. 
 Among the novelties compared to the six- and five-dimensional case is the appearance of $G_4$-flux as analysed in \cite{Mayrhofer:2014haa}.
 On $P_Q$ a $G_4$-flux of the form
 \bea \label{G4flux}
 G_4(P) = [\sigma_1] - \frac{1}{2} \, U_{\rm bi} \wedge P
 \eea
will in general compensate for the change in the Euler characteristic from $\hat P_Q$ to $P_Q$, analogously to the previously discussed conifold transitions in F/M-theory of \cite{Braun:2011zm,Krause:2012yh,Intriligator:2012ue}. 
In (\ref{G4flux})
\begin{equation}
\begin{aligned}
\sigma_0 &= \{ u = 0 \} \cap \{ w = 0 \}  \cap \{\rho = 0 \}, \\
\sigma_1 &= \{ u = 0 \} \cap \{ w = - b_2 v^2 \}  \cap \{\rho = 0 \}
\end{aligned}
\end{equation}
are four-cycles on $P_Q$ described as complete intersections in the ambient $\mathbb P_{112}$-fibration over $B_3$
and the flux $G_4$ fixes the complex structure such that $c_4 = \rho \,  \tau$ and $P: \{\rho =0\}$.\footnote{This assumes vanishing flux on $\hat P_Q$, see \cite{Mayrhofer:2014haa} for generalisations.}
Let us briefly describe the effect of this $G_4$ on the massless spectrum of the compactification due to M2-branes
wrapping the fibre components over the locus $C$.
In M-theory compactified on $P_Q$ to three dimensions
the charges of the states arising from M2-branes along $A_C$ and $B_C$ with respect to the surviving $U(1)_{S+U}$ can be read off from TABLE~\ref{hatPQcharges} by identifying $S+U$ on $\hat P_Q$ with $U_{\rm bi}$ on $P_Q$. Since states from $A_C$ and $B_C$ carry the same quantum numbers, counting the total number of charged zero-modes requires adding up the zero-mode excitations from both fibre components.
The integrals
\bea
\int_{A_C} G_4 = \frac{1}{2} \int_C P, \qquad \int_{B_C} G_4 = - \frac{1}{2} \int_C P
\eea
count the separate chiral index of states on $A_C$ and $B_C$. Note that both quantities are integer because the homology class of $C$ is even. 
Adding up both contributions we see that the net chirality with respect to $U(1)_{S+U}$ induced by the flux vanishes. 
In the four-dimensional  F-theory model on $P_Q$, $U(1)_{S+U}$ is replaced by the $\mathbb Z_2$ symmetry as discussed, and states along $A_C$ and $B_C$ both carry $\mathbb Z_2$ charge $1$ mod $2$. Clearly, there is no notion of chirality associated with this $\mathbb Z_2$, and this is well in agreement with the property of the flux of not inducing any $U(1)_{S+U}$ chirality already in M-theory.

\section{Torsion from the Weierstra\ss{} fibration} \label{sec:PW}
The second class of fibrations describing the same six-dimensional F-theory compactification is given by the Jacobian associated with the fibration $P_Q$ \cite{Braun:2014oya,Morrison:2014era}. It takes the form of a non-generic Weierstra\ss{} model
\bea
P_W = y^2 - x^3 - f x z^4 - g z^6 \label{jacwe}
\eea
with $[x:y:z]$ homogeneous coordinates of $\mathbb P_{231}$ and 
\bea \label{fg}
f &=&  e_1 \, e_3 - \frac{1}{3} e_2^2 - 4 e_0 \, e_4, \\
g &=& - e_0 e_3^2 + \frac{1}{3}  e_1 e_2 e_3 - \frac{2}{27} e_2^3 + \frac{8}{3} e_0 e_2 e_4 - e_1^2 e_4 \nonumber
\eea
where the $e_i$'s are given by
\bea \label{es}
e_0 &=& - c_0 + \frac{1}{4} b_0^2,  \qquad e_1 = - c_1 + \frac{1}{2} b_0 b_1, \nonumber  \\
e_2 &=&  - c_2 + \frac{1}{2} b_0 b_2 + \frac{1}{4} b_1^2,   \qquad e_3 = - c_3 + \frac{1}{2} b_1 b_2, \nonumber \\
e_4 &=& -c_4 + \frac{1}{4} b_2^2.
\eea

We focus again on a two-complex dimensional base space $B_2$.
While $P_Q$ and $P_W$ have the same discriminant, their fibre structure differs in two crucial ways \cite{Braun:2014oya}:
Unlike $P_Q$, the Weierstra\ss{} model
does have a holomorphic zero-section $Z: z=0$.
Second, $P_W$ exhibits 
non crepant-resolvable $I_2$-singularities over the specific locus $C$ on $B_2$ over which the fibre in $P_Q$ is a smooth $I_2$ fibre. 

The Weierstra\ss{} model $P_W$ is again related via a conifold transition to a smooth model $\hat P_W$. This resolved model can be identified with the geometry
of $\hat P_Q$ by mapping (blowing-up) $P_W|_{c_4 \equiv 0}$ to the Bl$^1\mathbb P_{112}[4]$-fibration over $B_2$.
The conifold transition occurs as the 2-step process
\bea
\hat P_W \rightarrow P_W|_{c_4 \equiv 0} \rightarrow P_W. \label{bdw}
\eea
As pointed out in \cite{Morrison:2012ei,Mayrhofer:2014haa}, the crucial difference compared to the transition relating $\hat P_Q$ to $P_Q$ is that now
in passing from $\hat P_W \rightarrow P_W|_{c_4 \equiv 0}$ the fibre component  ${B_I}$ and, simultaneously, $B_{II}$ shrink to zero size. Indeed the fibration structure and the intersection numbers in TABLE~\ref{hatPQcharges} allow us to deduce the K\"ahler cone on Bl$^1\mathbb P_{112}[4]$ relevant for the curves in the fibre. This  (part of the) K\"ahler form  is given by
\[ J=t_1\,U+t_2\,(S+U)\,\] with $t_1,\,t_2>0$. Integrating this two-form over the curves $A_{I}$, $A_{II}$, $B_I$ and $B_{II}$ yields
\[\int_{A_I}J=t_1,\, \int_{B_I}J=2\,t_2,\, \int_{A_{II}}J=t_1+t_2,\, \int_{B_{II}}J=t_2\,.\]
Therefore, we can identify the blow-down to the singular quartic (\ref{bdq}) with the limit $t_1 \rightarrow 0$, while the blow-down to the singular Weierstra\ss{} (\ref{bdw}) corresponds to $t_2 \rightarrow 0$. What becomes massless after this shrinking are M2-branes wrapping ${B_I}$ (and also those wrapping $B_{II}$). The M2-branes along the vanishing ${B_I}$ furnish the Higgs field
which acquires a VEV upon deforming the model from $P_W|_{c_4 \equiv 0} \rightarrow P_W$. The states associated with $B_{II}$ are mere spectators in this process.
From TABLE~\ref{hatPQcharges} we read off that e.g.\ under  $U(1)_U \times U(1)_{S-U}$ the Higgs field has charges $(0,2)$. Hence, there does not exist any unimodular transformation to a different basis in which the Higgs field is charged only under one of the $U(1)$s with charge 1. As a result it breaks 
\bea \label{U1breakingPW}
U(1)_U \times U(1)_{S-U} \rightarrow U(1)_U \times \mathbb Z_2 ,
\eea
in contrast to (\ref{U1breakingPQ}). 
Thus compactification of M-theory to five dimensions on $P_W$ does exhibit a bona fide $\mathbb Z_2$ symmetry. Since the Weierstra\ss{} model has a zero-section, standard duality to F-theory in six dimensions turns $U(1)_U$ into part of the six-dimensional diffeomorphism invariance and only the $\mathbb Z_2$ symmetry remains. As we have seen the mechanism how this $\mathbb Z_2$ comes about in F-theory on $P_W$ is very different to the $P_Q$ model.

This prompts the quest for torsional cohomology in the geometry $P_W$. 
To understand this we now analyze the conifold transition from the smooth $\hat P_W$ to $P_W$ in more detail. 
The conifold transition occurs along the lines of the well-known general analysis of \cite{Strominger:1995cz,Greene:1995hu,Greene:1996dh} except for some peculiarities which to the best of our knowledge have not been addressed before and which are responsible for the appearance of torsion. 

On $\hat P_W$ the locus $C_I = \{b_2 =0 \} \cap \{c_3=0\}$ consists of $N = [b_2] \cdot [c_3]$ points on the base $B_2$ of the fibration over which the fibre factorises. 
Let us label the two fibre components by $B_I^i$ and $A_I^i$ with $i=1,\ldots,N$. Due to the fibration structure all $B_I^i$ are homologous to each other. This gives rise to $N-M = N-1$ homology relations of the form
$B_I^1 = B_I^j$ for $j=2, \ldots, N$. Each of these homology relations is associated with a 3-chain $\Gamma_{1j}$ with $\partial \Gamma_{1j} = B_I^1 - B_I^j$.
The conifold transition first shrinks the $B_I^i$ to zero size and then deforms them into 3-spheres $S_3^i$. Following the general arguments of \cite{Strominger:1995cz,Greene:1995hu,Greene:1996dh}, the 3-spheres enjoy $M=1$ homology relations such that the number of independent spheres after the deformation is $N-1$.

Note once more that at the same time as the $B_I^i$ shrink, also the fibre component $B_{II}$ over the locus $C_{II}$ shrinks to zero size, but the deformation corresponding to switching on $c_4$ does not deform the resulting singularities into 3-spheres. This is just the statement that on $P_W$ non-crepant resolvable $I_2$ loci in the fibre remain. We will return to the fate of these singularities later.

\begin{figure*}
\begin{tikzpicture}[>=stealth]
\begin{scope}[xshift=-5cm,yshift=-1.5cm,rotate=90]
\shadedraw[xshift=0.4cm,rotate=180,ball color=blue!40!white] plot [smooth , tension=0.9] coordinates {(-1,2) (-.8,.8) (-0.4,0.4) (0,0) (-.4,-.4) (-.8,-.8)(-1,-2) };
\shadedraw[ball color=blue!40!white] plot [smooth cycle, tension=0.9] coordinates {(0,0) (0.8,1.2) (1.5,0.8) (1,0) (1.5,-0.8) (0.8,-1.2) };
\draw[green!50!white, line width=0.5mm, xshift=.5cm,yshift=.5cm] (-0.1,-0.1)--(0.1,0.1);
\draw[green!50!white, line width=0.5mm, xshift=.5cm,yshift=.5cm] (-0.1,0.1)--(0.1,-0.1);
\begin{scope}[xshift=3cm,xscale=-1]
\shadedraw[ball color=gray!30!white] plot [smooth cycle, tension=0.9] coordinates {(0,0) (0.8,1.2) (1.5,0.8) (1,0) (1.5,-0.8) (0.8,-1.2) };
\end{scope}
\shadedraw[xshift=.5cm,yshift=.5cm,ball color=green!50!white] plot [smooth , tension=0.9] coordinates {(-1,1.5) (-.8,.8) (-0.4,0.2) (0,0) (-.4,-.2) (-.8,-.8)(-1,-1.5) };
\end{scope}
\draw[->] (-2.8,0)--(-.3,0) node [midway,above] {blow down} 
                node [midway,below] {to sing.\ Weierstra\ss{}};
\begin{scope}[xshift=1.5cm,yshift=-1cm,rotate=-90,xscale=-1]
\shadedraw[xshift=.5cm,,rotate=180,ball color=blue!40!white] plot [smooth , tension=0.9] coordinates {(-1,2) (-.8,.8) (-0.4,0.2) (0,0) (-.4,-.2) (-.8,-.8)(-1,-2) };
\shadedraw[ball color=blue!30!white] (0,0) to[out=90,in=180] (1.2,1.2) to[out=0,in=70] (2.0,0)
            to[out=110,in=0] (1.4,0.5) to[out=180,in=90] (1,0) to[out=-90,in=180] (1.4,-0.5)
            to[out=0,in=-110] (2,0) to[out=-70,in=0] (1.2,-1.2) to[out=180,in=-90] (0,0);
\draw[green!50!white, line width=0.5mm, xshift=.8cm,yshift=-5mm] (-0.1,-0.1)--(0.1,0.1) (-0.1,0.1)--(0.1,-0.1);
\shadedraw[xshift=.8cm,yshift=-5mm,ball color=green!50!white,rotate=0] plot [smooth , tension=0.9] coordinates {(-1,1.5) (-.8,.8) (-0.4,0.2) (0,0) (-.4,-.2) (-.8,-.8)(-1,-1.5) };
\end{scope}
\draw[->] (4,0)--(6.5,0) node [midway,above] {deformation} 
                node [midway,below] {of sing.\ Weierstra\ss{}};
\begin{scope}[xshift=8cm]
\shadedraw[ball color=blue!30!white] (-2,1) to [bend left] (-.5,0)  arc (0:-180:-.5 and 0.2) to [bend left] (1.5,0) arc (0:-180:-.5 and .2) to [bend left] (4,1);
\draw[decoration={markings, mark=at position 0.3 with {\arrow{>}}, mark=at position 0.8 with {\arrow{>}}},
      postaction={decorate},thick]
      (-.5,0)  arc (0:-180:-.5 and 0.2)
      (1.5,0) arc (0:-180:-.5 and 0.2);
\draw[dashed,decoration={markings, mark=at position 0.3 with {\arrow{<}}, mark=at position 0.8 with {\arrow{<}}},
      postaction={decorate},thick]
      (-.5,0)  arc (0:180:-.5 and 0.2)
      (1.5,0)  arc (0:180:-.5 and 0.2);
\draw[green!50!white, line width=0.5mm, xshift=.8cm,yshift=3mm] (-0.1,-0.1)--(0.1,0.1) (-0.1,0.1)--(0.1,-0.1);
\shadedraw[xshift=.8cm,yshift=3mm,ball color=green!50!white,rotate=90] plot [smooth , tension=0.9] coordinates {(-1,1.5) (-.8,.8) (-0.4,0.2) (0,0) (-.4,-.2) (-.8,-.8)(-1,-1.5) };
\end{scope}
\end{tikzpicture}
\caption{Figure showing the boundaries induced after the conifold transition in the Weierstra\ss{} hypersurface in $\mathbb P_{231}$. The divisor $S$ is denoted in blue and $U$ is denoted in green. After the transition $U$ does not develop a boundary and therefore is associated to the five-dimensional $U(1)_U$ symmetry. On the other hand $S$ develops two boundaries of the same orientation. The sum over all the points $B^i_{I}$ for {\it each one} of the two boundaries illustrated gives the torsional 3-cycle associated to the $\mathbb{Z}_2$ symmetry.}\label{fig:conifold-transition-Weierstrass-case}
\end{figure*}
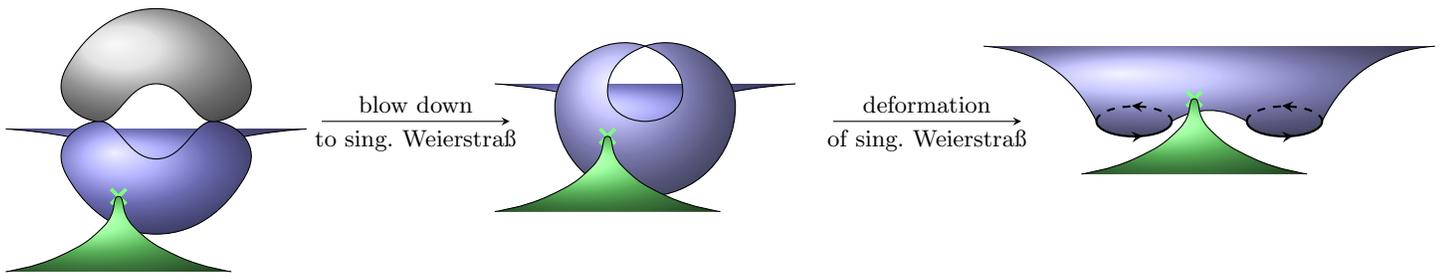

According to the general analysis of the conifold transition with $M=1$, there must exist one `magnetic' 4-cycle $D$ which intersects each of the two-spheres $B_I^i$. This 4-cycle is given by the divisor $S$ with intersection numbers
\bea \label{DAI}
D \cdot B_I^i = 2.
\eea
Indeed, the rational section $S$ wraps the entire fibre $A_I^i$, and the two intersection points with $B_I^i$ are evident from FIG.~\ref{fig:fibre}. Importantly, the other section $U$ does not intersect the $B_I^i$ and therefore, since the $B_I^i$ are fibral curves, the fibration structure guarantees that no other integer four-cycle exists intersecting the $B_I^i$. In particular there exists no such divisor with intersection number $1$. 
After shrinking the $B_I^i$ cycles to nodes and deforming them into $S_3^i$, each one induces a boundary on $S$ turning it into a 4-chain. The crucial peculiarity of the conifold transition $\hat P_W \rightarrow P_W$ is that $S$ intersects the two-spheres at \emph{two} points and so they each induce a boundary of the same orientation. Thus the precise homological relation obeyed by the $S_3^i$ is
\bea
2 \,  \Gamma = \partial \hat D, \qquad \Gamma = \sum_i S_3^i.
\eea
This is illustrated in FIG.~\ref{fig:conifold-transition-Weierstrass-case}.
We therefore identify the 3-cycle $\Gamma$ as a $\mathbb Z_2$ element of ${\rm Tor} H_3(P_W, \mathbb Z)$. For a generic base $B_2$ this is the only such element and so
\bea
{\rm Tor} H_3(P_W, \mathbb Z) = \mathbb Z_2.
\eea
The argument about the boundary of the 4-chain $\hat D$ after the transition made crucial use of the fact that $D$ intersects each of the shrinking 2-cycles $B_I^i$. Note that in addition, $D$ also intersects the shrinking fibre component $B_{II}$ over $C_{II}$ as is evident from TABLE~\ref{hatPQcharges}. 
As will be discussed in more detail momentarily, it is possible to resolve these singularities after a suitable blow-up in the base $B_2$ as in \cite{DolgacheGross}. This will replace the former intersection points with $D$ by an even-dimensional cycle and thus does not induce any additional boundaries for the 4-chain $\hat D$ which could spoil the argument. Consistently, the general analysis of \cite{DolgacheGross} shows that after the blow-up in the base and resolving, the resulting geometry possesses non-trivial torsional cohomology. 

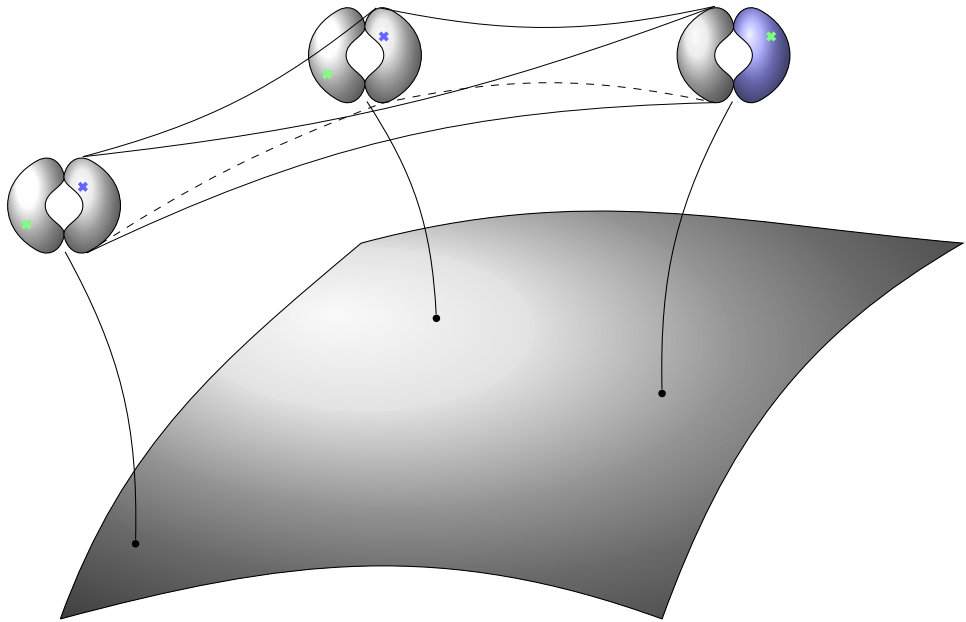
\begin{figure*}
\begin{tikzpicture}[>=stealth]
\shadedraw[ball color=gray!30!white] (0,0) to[out=15,in=160]
(8,0) to[out=70,in=-150]
(12,5) to[out=175,in=15]
(4,5) to[out=-140,in=70] (0,0);
\node[fill,circle, minimum size=1mm,inner sep=0] (bII1) at (1,1) {};
\node (fII1) at (0,5) {};
\draw[thin] (bII1) to[bend right=15] (fII1);
\node[fill,circle, minimum size=1mm,inner sep=0] (bII2) at (5,4) {};
\node (fII2) at (4,7) {};
\draw[thin] (bII2) to[bend right=15] (fII2);
\node[fill,circle, minimum size=1mm,inner sep=0] (bI1) at (8,3) {};
\node (fI1) at (9,7) {};
\draw[thin] (bI1) to[bend left=15] (fI1);
\begin{scope}[xshift=9.7cm,yshift=7.5cm,scale=-.5]
\shadedraw[ball color=blue!40!white] plot [smooth cycle, tension=0.9] coordinates {(0,0) (0.8,1.2) (1.5,0.8) (1,0) (1.5,-0.8) (0.8,-1.2) };
\draw[green!50!white, line width=0.5mm, xshift=.5cm,yshift=-5mm] (-0.1,-0.1)--(0.1,0.1) (-0.1,0.1)--(0.1,-0.1);
\shadedraw[ball color=gray!20,xscale=-1,xshift=-3cm] plot [smooth cycle, tension=0.9] coordinates {(0,0) (0.8,1.2) (1.5,0.8) (1,0) (1.5,-0.8) (0.8,-1.2) };
\end{scope}
\begin{scope}[xshift=3.3cm,yshift=7.5cm,scale=.5]
\shadedraw[ball color=gray!1] plot [smooth cycle, tension=0.9] coordinates {(0,0) (0.8,1.2) (1.5,0.8) (1,0) (1.5,-0.8) (0.8,-1.2) };
\draw[green!50!white, line width=0.5mm, xshift=.5cm,yshift=-5mm] (-0.1,-0.1)--(0.1,0.1) (-0.1,0.1)--(0.1,-0.1);
\shadedraw[ball color=gray!20,xscale=-1,xshift=-3cm] plot [smooth cycle, tension=0.9] coordinates {(0,0) (0.8,1.2) (1.5,0.8) (1,0) (1.5,-0.8) (0.8,-1.2) };
\draw[blue!60!white, line width=0.5mm, xshift=2cm,yshift=5mm] (-0.1,-0.1)--(0.1,0.1) (-0.1,0.1)--(0.1,-0.1);
\end{scope}
\begin{scope}[xshift=-.7cm,yshift=5.5cm,scale=.5]
\shadedraw[ball color=gray!1] plot [smooth cycle, tension=0.9] coordinates {(0,0) (0.8,1.2) (1.5,0.8) (1,0) (1.5,-0.8) (0.8,-1.2) };
\draw[green!50!white, line width=0.5mm, xshift=.5cm,yshift=-5mm] (-0.1,-0.1)--(0.1,0.1) (-0.1,0.1)--(0.1,-0.1);
\shadedraw[ball color=gray!20,xscale=-1,xshift=-3cm] plot [smooth cycle, tension=0.9] coordinates {(0,0) (0.8,1.2) (1.5,0.8) (1,0) (1.5,-0.8) (0.8,-1.2) };
\draw[blue!60!white, line width=0.5mm, xshift=2cm,yshift=5mm] (-0.1,-0.1)--(0.1,0.1) (-0.1,0.1)--(0.1,-0.1);
\end{scope}
\draw[very thin] (0.35,4.87) to[bend left=12] (8.7,6.87);
\draw[very thin] (0.3,6.15) to[bend right=12] (4.2,8.13);
\draw[very thin] (4.25,8.13) to[bend right=12] (8.7,8.15);
\draw[very thin] (0.3,6.15) to[bend right=7] (8.7,8.15);
\draw[very thin,dashed] (4.27,6.86) to[bend left=12] (8.7,6.87);
\draw[very thin,dashed] (0.35,4.87) to[bend left=6] (4.27,6.86);

\end{tikzpicture}
\caption{Figure showing the 3-chains stretching between a point in the set of points $C_I$ and two points in the set $C_{II}$ in the resolved space. The boundary of the chain is therefore $2 B_{II} - B_{I}$. After the deformation the boundary $B_{I}$ is lost leaving a chain with a boundary $2 B_{II}$ and thereby identifying $B_{II}$ as the torsional 2-cycle.}\label{fig:chain-relation-B}
\end{figure*}

Having identified a non-trivial $\mathbb Z_2$ element in ${\rm Tor} H_3(P_W, \mathbb Z)$ the universal coefficient theorem implies that on a smooth manifold also ${\rm Tor} H_2(P_W, \mathbb Z)$ is non-trivial. In order to identify these torsional cycles consider the fully resolved  ${\rm Bl}^1 \mathbb P_{112}[4]$-fibration. We are interested in the homology classes of the fibre components $A_I$, $B_I$, $A_{II}$ and $B_{II}$. Since there are only two homologically independent sections these four fibre components must enjoy certain homology relations. Being fibral curves they only intersect the sections $S$ and $U$ (in the absence of non-abelian gauge symmetries) and so these intersection numbers, as given in TABLE~\ref{hatPQcharges}, determine uniquely their homology classes. In particular we see that in homology $2 B_{II} = B_{I}$, which means that there are 3-chains stretching between a point in the set of points $C_I$ and two points in the set $C_{II}$ with a boundary $2 B_{II} - B_{I}$. See FIG.~\ref{fig:chain-relation-B} for an illustration of this.
Now as we perform the conifold transition over the $C_I$ loci the $B_I$ shrink and then are deformed as $S^3$s and so no longer form boundaries to these 3-chains. If we were able to perform these transitions over the $C_I$ loci without affecting the $C_{II}$ loci the remaining 3-chains would have a boundary $2 B_{II}$ and so the $B_{II}$ would be associated to the expected torsional 2-cycles. This is essentially the correct identification, however there is a subtlety due to the fact that necessarily the $B_{II}$ must simultaneously shrink in order to be able to perform the deformation.

In order to see the torsional cycles on a smooth manifold we must resolve the loci $C_{II}$. First we note that since we have found the torsional 3-cycles explicitly, any smooth resolution implies the existence of the torsional 2-cycles identified above via the universal coefficient theorem. One way to perform the resolution is by a small resolution which would lead to a non-K\"ahler manifold, but as stated would be sufficient to identify the torsional cycles (see \cite{Aspinwall:1995rb} for examples of this process). Another way, following \cite{DolgacheGross}, is by blowing up the base over the $C_{II}$ locus and then resolving the resulting $SU(2)$ singularity over the exceptional divisor of this base blow-up ${T}_{II}$. The resulting space is not Calabi-Yau but still K\"ahler. Over certain points $\hat{C}_{II}$ on $T_{II}$ the fibre will enhance to type $I_3$ (corresponding to $SU(2)$-matter). In this smooth geometry the torsional 2-cycles can be identified as before in terms of 3-chains stretching between the points $C_I$ and $\hat{C}_{II}$. In the appendix we present a detailed analysis of this procedure and in particular identify the explicit components of the fibre which after the deformation become the torsional 2-cycles. 

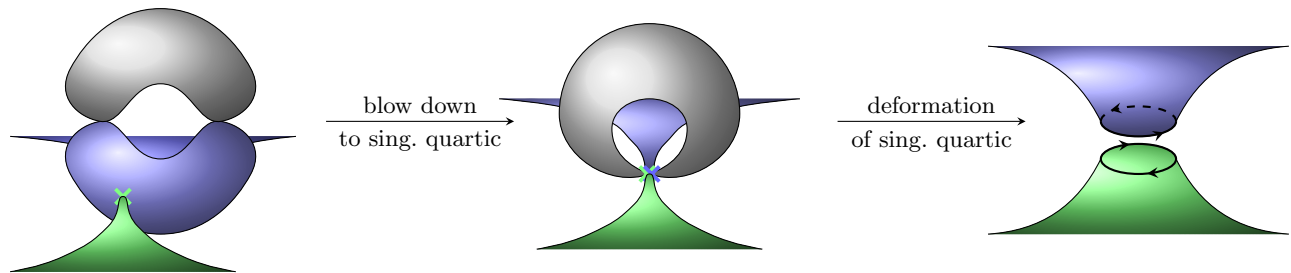
\begin{figure*}
\begin{tikzpicture}[>=stealth]
\begin{scope}[xshift=-5cm,yshift=-1.5cm,rotate=90]
\shadedraw[xshift=0.3cm,rotate=180,ball color=blue!50!white] plot [smooth , tension=0.9] coordinates {(-1,1.8) (-.8,.8) (-0.4,0.4) (0,0) (-.4,-.4) (-.8,-.8)(-1,-2) };
\shadedraw[ball color=blue!40!white] plot [smooth cycle, tension=0.9] coordinates {(0,0) (0.8,1.2) (1.5,0.8) (1,0) (1.5,-0.8) (0.8,-1.2) };
\draw[green!50!white, line width=0.5mm, xshift=.5cm,yshift=.5cm] (-0.1,-0.1)--(0.1,0.1);
\draw[green!50!white, line width=0.5mm, xshift=.5cm,yshift=.5cm] (-0.1,0.1)--(0.1,-0.1);
\begin{scope}[xshift=3cm,xscale=-1]
\shadedraw[ball color=gray!30!white] plot [smooth cycle, tension=0.9] coordinates {(0,0) (0.8,1.2) (1.5,0.8) (1,0) (1.5,-0.8) (0.8,-1.2) };
\end{scope}
\shadedraw[xshift=.5cm,yshift=.5cm,ball color=green!50!white] plot [smooth , tension=0.9] coordinates {(-1,1.5) (-.8,.8) (-0.4,0.2) (0,0) (-.4,-.2) (-.8,-.8)(-1,-1.5) };
\end{scope}
\draw[->] (-2.8,0)--(-.3,0) node [midway,above] {blow down} 
                node [midway,below] {to sing.\ quartic};
\begin{scope}[xshift=1.5cm,yshift=1.3cm,rotate=90,xscale=-1]
\shadedraw[xshift=2cm,rotate=0,ball color=blue!40!white] plot [smooth , tension=0.9] coordinates {(-1,2) (-.8,.8) (-0.4,0.2) (0,0) (-.4,-.2) (-.8,-.8)(-1,-2) };
\shadedraw[ball color=gray!30!white] (0,0) to[out=90,in=180] (1.2,1.2) to[out=0,in=70] (2.0,0)
            to[out=110,in=0] (1.4,0.5) to[out=180,in=90] (1,0) to[out=-90,in=180] (1.4,-0.5)
            to[out=0,in=-110] (2,0) to[out=-70,in=0] (1.2,-1.2) to[out=180,in=-90] (0,0);
\draw[green!50!white, line width=0.5mm, xshift=2cm,yshift=0.03cm] (-0.1,-0.1)--(0.1,0.1) (-0.1,0.1)--(0.1,-0.1);
\draw[blue!60!white, line width=0.5mm, xshift=2cm,yshift=-.03cm] (-0.1,-0.1)--(0.1,0.1) (-0.1,0.1)--(0.1,-0.1);
\shadedraw[xshift=2cm,ball color=green!50!white,rotate=180] plot [smooth , tension=0.9] coordinates {(-1,1.5) (-.8,.8) (-0.4,0.2) (0,0) (-.4,-.2) (-.8,-.8)(-1,-1.5) };
\end{scope}
\draw[->] (4,0)--(6.5,0) node [midway,above] {deformation} 
                node [midway,below] {of sing.\ quartic};
\begin{scope}[xshift=8cm]
\shadedraw[ball color=blue!40!white] (-2,1) to [bend left] (-.5,0)  arc (0:-180:-.5 and 0.2) to [bend left] (2,1);
\draw[decoration={markings, mark=at position 0.8 with {\arrow{>}}},
      postaction={decorate},thick]
      (-.5,0)  arc (0:-180:-.5 and 0.2);
\draw[dashed,decoration={markings, mark=at position 0.3 with {\arrow{<}}},
      postaction={decorate},thick]
      (-.5,0)  arc (0:180:-.5 and 0.2);
\end{scope}
\begin{scope}[xshift=8cm,yshift=-.5cm,yscale=-1]
\shadedraw[ball color=green!50!white] (-2,1) to [bend left] (-.5,0)  arc (0:-180:-.5 and 0.2) to [bend left] (2,1);
\draw[decoration={markings, mark=at position 0.2 with {\arrow{>}}, mark=at position 0.7 with {\arrow{>}}},
      postaction={decorate},thick]
      (0,0) ellipse (.5 and .2);
\end{scope}
\end{tikzpicture}
\caption{Figure showing the boundaries induced after the conifold transition in the quartic hypersurface in $\mathbb P_{112}$. The divisors $S$, denoted in blue, and $U$, denoted in green, both develop a single boundary from each of the $A_I^i$ of opposite orientation. The two boundaries are then glued together to form the divisor $S+U$ corresponding to the remnant five-dimensional $U(1)$ symmetry. }\label{fig:conifold-transition-P112-case}
\end{figure*}

With this understanding it is worth returning to the $P_Q$ fibration to see why this torsion is absent there. Now the divisors $S$ and $U$ both develop a single boundary from each of the $A_I^i$ of opposite orientation. Therefore we can make a cycle from these two chains by gluing these boundaries forming the divisor $S+U$. The other four-chain associated to $U$ has only a single boundary and therefore no torsion element arises. This is illustrated in FIG.~\ref{fig:conifold-transition-P112-case}.
Note that if we chose to consider the 4-chain corresponding to $S-U$ it would indeed have 2 boundaries of the same orientation at each locus, but this would not imply torsion since there are other 4-chains, $S$ and $U$, which have half the boundary of $S-U$. Consistent with this we see that the identification of the torsional 2-cycles presented in the above paragraphs is also modified because now it is the $A_I$ components which shrink and are removed as boundaries to 3-chains associated to homology relations between the components. But this does not create a new torsion cycle but just implies in homology that $A_{II}=B_{II}$.

By Poincar\'e duality the existence of a $\mathbb Z_2$ torsional 3-cycle implies the existence of a 3-form $\alpha$ such that
\bea
2 \,  \alpha = d \tw.
\eea
The non-closed 2-form $\tw$ is the Poincar\'e dual to the 4-chain $\hat D$ and can be interpreted as the generator of the $\mathbb Z_2$ symmetry. Expanding the M-theory 3-form $C_3$ as $C_3 = A \wedge \tw + \ldots$ gives rise to a massive $U(1)$ gauge field in five spacetime dimensions which precisely corresponds to the original $U(1)_{S-U}$ gauge symmetry after Higgsing (see \cite{Grimm:2011tb} for a discussion of this mechanism in the context of F-theory). 

We conclude by discussing the physical significance of the identified torsional cycles.\footnote{See \cite{Camara:2011jg} for an analogous analysis in four-dimensional Type II compactifications.}
As alluded to already in the introduction, a $\mathbb Z_2$ gauge theory contains a characteristic set of Wilson line operators \cite{Banks:2010zn}. If the theory contains physical $\mathbb Z_2$ electrically charge particles, these operators can be interpreted as describing the associated world-line. However, from a quantum field theoretic perspective  the Wilson line operators exist even  in absence of any $\mathbb Z_2$ charged particle in the physical particle spectrum. In quantum gravity, by contrast, it is conjectured that the full lattice of possible charges is populated by (possibly massive) physical states \cite{Banks:2010zn}.
To appreciate how this conjecture is indeed confirmed in our M/F-theoretic setting,
consider the five-dimensional effective field theory associated with M-theory compactified on $P_W$. The Wilson line operators describe the word-line of M2-branes wrapping the identified torsional 2-cycles, which do exist as physical particles, in perfect agreement with the above conjecture. One might wonder if a modification of the geometry would be possible that gives rise to a $\mathbb Z_2$ gauge theory without such physical $\mathbb Z_2$ charged particles. As we have seen, the torsional 2-cycles wrapped by the associated M2-branes are related to the fibre components over $C_{II}$ before the deformation. The class of $C_{II}$ depends on the class of the coefficients $c_i$ and $b_i$ defining the Weierstra\ss{} model. 
Recall that these transform as sections of certain line bundles on the base.
One might try to exploit the existing freedom in choosing these line bundles  to arrange for the cohomology class of the locus $C_{II}$ to be trivial, in which case no $\mathbb Z_2$ charged states would exist. It is easy to see by direct inspection of the coefficient classes (cf.\ e.g.\ Table~1 of \cite{Mayrhofer:2014haa}), however, that this also removes the Higgs field along $C_{I}$ and thus destroys the $\mathbb Z_2$ gauge theory in the first place. This is of course in agreement with the universal coefficient theorem which guarantees that ${\rm Tor} H_3(P_W, \mathbb Z) \simeq {\rm Tor} H_2(P_W, \mathbb Z)$. 

In five dimensions, the magnetic dual to an electrically charged particle is a string. In our setting these magnetic objects again exist as physical objects arising from M5-branes wrapping the 4-chain $\hat D$. Since $\hat D$ has the boundary $2 \,  \Gamma$, one can consider 
a configuration consisting of an M5-brane on the 4-chain $\hat D$ together with two M5-branes on $\Gamma$. This is the M-theory analogue of the configuration considered before in \cite{Greene:1996dh} with the important difference that here the M5-branes on the boundary of $\hat D$ give rise to \emph{two} membranes in five dimensions ending on the (`magnetic') string.
This again realises the expectations based on the general framework of $\mathbb Z_2$ gauge theory described in \cite{Banks:2010zn}: In a four-dimensional $\mathbb Z_k$ gauge theory, $k$ units of flux tubes (strings) end on a magnetic monopole  to turn the full configuration into a stable object, and in five dimensions the strings and magnetic monopoles become membranes and `magnetic' strings.
\\

\section{Conclusions}

In this paper we have studied the relation between discrete gauge symmetry and torsional homology in F/M-theory. 
A $\mathbb Z_k$ discrete symmetry in F-theory is associated with $k$ isomorphism classes of inequivalent genus-one fibrations with the same Jacobian. These form the Tate-Shafarevich group associated with the Jacobian.
Focusing here on $k=2$ for simplicity we have studied in detail the two different fibrations $P_Q$ and $P_W$ associated with the appearance of a $\mathbb Z_2$ discrete gauge group in F-theory, where $P_Q$ is a smooth $\mathbb P_{112}[4]$-fibration and $P_W$ represents its singular Jacobian fibration in Weierstrass form \cite{Braun:2014oya}.
Compactifications on $P_Q$ have been studied in quite some detail recently in \cite{Braun:2014oya,Morrison:2014era,Anderson:2014yva,Klevers:2014bqa,Garcia-Etxebarria:2014qua,Mayrhofer:2014haa}.
We have shown in this article that compactification of M-theory on $P_Q$ gives rise to a fibral $U(1)$ gauge symmetry and discussed how this symmetry is lifted  to a $\mathbb Z_2$ symmetry in the F-theory limit. By contrast, M-theory on $P_W$ yields fibral gauge group $U(1) \times \mathbb Z_2$, of which only the $\mathbb Z_2$ part survives in the F-theory limit. 
Consistently with field theoretic expectations based on the different M-theory compactifications on $P_Q$ and $P_W$, it is only on $P_W$ that torsional homology arises. We have explicitly identified $\mathbb Z_2$ torsional 2- and 3-cycles by analyzing a birational blowup-up of $P_W$. On $P_Q$, on the other hand, torsional homology appears only in the formal sense of a $\mathbb Z_2$ torsional fibral 2-cycle {\emph{modulo the fiber class}}.  It would be interesting to explicitly generalize our analysis of the appearance of torsion to fibrations which give rise to higher discrete symmetry groups.

A rather different notion of torsion arises in the context of the Mordell-Weil group of rational sections. The Mordell-Weil group can have not only a free part corresponding to massless $U(1)$ symmetries, but also a torsional component. 
As conjectured in \cite{Aspinwall:1998xj} and proven in \cite{Mayrhofer:2014opa}, such Mordell-Weil torsion enforces non-simply connected non-abelian gauge groups in F-theory. 
Geometrically, fibrations with torsional Mordell-Weil group exhibit torsional divisors {\emph{modulo the Cartan divisors} associated with (parts of) the non-abelian gauge group \cite{Mayrhofer:2014opa}.
Interestingly, fibrations with Mordell-Weil torsion and with discrete gauge symmetries are interchanged by mirror symmetry in the fiber \cite{Klevers:2014bqa}. This suggests an intriguing connection between the Tate-Shafarevich group underlying discrete symmetries and the torsion component of the Mordell-Weil group of rational sections, which will be exciting to further study in more detail.

\section*{Acknowledgements}

We thank Florent Baume, I\~naki Garc\'i{}a-Etxebarria, Luca Martucci, and in particular Arthur Hebecker for very useful discussions.
We are grateful to MITP Mainz for hospitality during the workshop `String Theory and its Applications', where part of this work was carried out. 
%
%

\appendix
\hypertarget{sec:blow-up}{\section{Blowing up the Matter Locus}}

As outlined in the main text, one way to identify the torsional 2-cycles in a smooth geometry is by blowing up the $C_{II}$ locus in the base of the fibration \cite{DolgacheGross}. The resulting space is birationally equivalent to the original one and therefore allows one to deduce the torsional cohomology also for the latter \cite{Braun:2014oya}. In this appendix we give the technical details of this procedure. 

Let us begin with a simple example which we will build up to the final result. Consider the $U(1)$-restricted Tate model presented in \cite{Grimm:2010ez} given by the hypersurface $P_T$ in $\mathbb P_{231}$
\be
P_T = y^2 + a_1 x y z + a_3 y z^3 - x^3 - a_2 x^2 z^2 - a_4 x z^4= 0 \;. \label{u1restricted}
\ee
This fibration is a specialization of the Weierstrass model (\ref{jacwe}), (\ref{fg}), (\ref{es}) with no double-charged singlets. It exhibits two independent sections and a set of points with conifold singularities $T_{II}$ where matter with charge one with respect to the associated $U(1)$ symmetry  resides,
\be
T_{II} \;:\; a_3=a_4=0 \;.
\ee
One can resolve these singularities through a blow-up in the ambient variety, involving the fibre co-ordinates by sending $\left(x,y\right) \rightarrow \left(x s, y s \right)$ and imposing the scaling relation $\left(x,y,s\right) \sim \left(\lambda^{-1} x,\lambda^{-1} y,\lambda s \right)$. The blowup divisor $S:\;s=0$ acts as a rational section, in addition to the zero section $Z:\;z=0$. The resulting manifold is smooth and over the locus $a_3=a_4=0$ the fibre is of type $I_2$. 

Let us now consider starting from the singular fibration but instead of resolving we blow up the base over the locus $T_{II}$ by sending $\left(a_3,a_4\right) \rightarrow \left(a_3 t, a_4 t \right)$ and introducing the relation $\left(a_3,a_4,t\right) \sim \left(\lambda^{-1} a_3,\lambda^{-1} a_4,\lambda t \right)$. The resulting geometry now has an $SU(2)$ singularity over the exceptional divisor 
\bea
T\;:\;t=0.
\eea
 Note that after this replacement $t$ does not factor from $P_T$, which means that the proper transform of the hypersurface equation has non-vanishing fist Chern class, i.e.~ is not Calabi-Yau any more. Nonetheless the space is K\"ahler and we can proceed, though dynamically this configuration is unlikely to be stable due to the absence of supersymmetry. It merely serves as a birational auxiliary geometry which allows us to identify the torsional cycles. We can resolve the $SU(2)$ singularity in the standard way of resolving non-abelian singularities over divisors by performing a second blow-up involving now the fibre coordinates $\left(x,y,t\right) \rightarrow \left(x s, y s, t s \right)$ and identifying $\left(x,y,t,s\right) \sim \left(\lambda^{-1} x,\lambda^{-1} y,\lambda^{-1} t,\lambda s \right)$. The resolved fibration takes the form
\be
\hat{P}_T = y^2 + a_1 x y z + a_3 t y z^3 - s x^3 - a_2 x^2 z^2 - a_4 t x z^4= 0 \;. \label{u1restrictedblowup}
\ee
This is a smooth space. The fibre over a generic point on $T$ is of type $I_2$ with the two components 
\begin{equation}
\begin{aligned}
A_{II}:&\;T \cap \hat{P}_{T} \cap \left\{C_{\mathrm{base}}\right\} \;, \\
B_{II}:&\;S \cap \hat{P}_{T}\cap \left\{C_{\mathrm{base}}\right\} \;, \label{su2comp}
\end{aligned}
\end{equation}
where $\left\{C_{\mathrm{base}}\right\}$ is some curve in the base intersecting intersecting $T$ at a generic point. These two components intersect at two points.  

Over two sets of special points along the divisor $t=0$ in the base the fibre changes. The first set $D_{II}$ corresponds to the locus $D_{II}:\; \left\{t=0\right\} \cap \left\{ 4 a_2 + a_1^2 = 0\right\}$. Over this locus the fibre becomes of type $III$. There is no symmetry enhancement or matter states associated to this locus. The second more interesting locus is given by $\hat{C}_{II}:\;\left\{t=0\right\} \cap \left\{ a_2 a_3^2 - a_1 a_3 a_4 - a_4^2 = 0\right\}$. Over this locus of points the $B$ components of the fibre splits into 2 components
\be
\left. B_{II} \right|_{\hat{C}_{II}} \rightarrow B_{II,1} + B_{II,2}.
\ee
The fibre becomes type $I_3$ which signals the presence of matter transforming in the fundamental of $SU(2)$. 

There are two important ways that this toy example differs from the singular Weierstra\ss{} model we are interested in. The first, quantitative, difference is that the matter point locus $T_{II}$ in the example is very simple while the corresponding locus $C_{II}$ in the full model (\ref{jacwe}), (\ref{fg}), (\ref{es}) is very complicated. This makes performing the blow-up in the base, though conceptually equivalent, technically difficult. We will return to this later. The second, qualitative, difference is that in the full model there are two rather than one matter loci, $C_{I}$ and $C_{II}$. We can proceed by blowing up $C_{II} \rightarrow T$ as in the example above. However the key point is that the resolution $\left(x,y,t\right) \rightarrow \left(x s, y s, t s \right)$ will only resolve the $SU(2)$ singularity over $T$ but not the singularity of $C_{I}$. 

We will therefore require a further resolution. Importantly this will introduce another independent homology class for the components of the fibre independent of the Cartan of the $SU(2)$. Therefore now in the K\"ahler cone we will have an additional degeneration possibility where the $C_{I}$ locus becomes singular while the $T$ divisor remains smooth. In this limit we can then deform the $C_{I}$ locus and reach the smooth geometry with the $\mathbb{Z}_2$ discrete symmetry and torsion. Alternatively we can perform the deformation first and then blow up the base in the deformed model, since the blow-up is localised away from the deformation locus this should lead to the same result.

In the main text we have identified the torsional 2-cycles by studying the intersection numbers of the sections with the resolved Weierstra\ss{} model. This is equivalent to looking at their $U(1)$ charges. The intersection of the section with the components of the fibre over the $C_{I}$ locus remain unchanged by a blow-up in the base over the $C_{II}$ locus. Indeed it is clear that the component $B_I$ of the fibre over $C_{I}$ which shrinks and is then deformed must have vanishing intersection with $U$, since this remains as the zero section after the deformation; furthermore since the intersection with $S-U$ is the 6-dimensional $U(1)$ charge (of the massless Higgs), it is independent
 of the resolution. Therefore it must be that the shrinking component intersects $S$ with $+2$ and so the argument for the existence of the 3-chains goes through for the blown-up base geometry as long as we can identify components of the fibre which have the same intersection numbers as $B_{II}$ in TABLE~\ref{hatPQcharges}. Since this would mean they cannot intersect the Cartan of the $SU(2)$ they can only arise as combinations of the fibre components over the analogue of the matter points $\hat{C}_{II}$ in the full Weierstrass model $P_W$. They therefore will induce the 3-chains as described in the main text. 

Let us now turn to applying this procedure to the full model $P_W$ (\ref{jacwe}). As analysed in \cite{Morrison:2014era,Klevers:2014bqa,Mayrhofer:2014haa}, the single-charged locus $C_{II}$ is given by a complicated prime ideal. We shall use the particular form given in \cite{Morrison:2014era} where it is given by the (non-transversal) intersection of the 7 polynomials
\begin{equation}
\begin{aligned}
H_1 =& e_1 b^4 - 2 e_2 e_3 b^2 + 2 e_3^3,  \\
H_2 =& 2 e_0 b^4 - 2 e_2^2 b^2 + e_1 e_3 b^2 + 2 e_2 e_3^2,  \\
H_3 =& -e_1 e_2 b^2 + 2 e_0 e_3 b^2 + e_1 e_3^2,  \\
H_4 =& -e_1^2 b^2 + 4 e_0 e_3^2,  \\
H_5 =& 2 e_0 e_1 b^2 + e_1^2 e_3 - 4 e_0 e_2 e_3, \\
H_6 =& 4 e_0^2 b^2 + e_1^2 e_2 - 4 e_0 e_2^2 + 2 e_0 e_1 e_3, \\
H_7 =& e_1^3 - 4 e_0 e_1 e_2 + 8 e_0^2 e_3 \;.
\end{aligned}
\end{equation}
Here the $e_i$ are as in (\ref{fg}) and we are working with the singular geometry corresponding to $c_4 \equiv 0$, which implies $e_4=\frac{1}{4}b^2$ (after relabeling $b_2 \rightarrow b$). 
To blow up the zero-locus of this ideal we can introduce new coordinates $f_i$ and $t$ and write the blown-up space as the variety corresponding to the vanishing locus of the  ideal 
\be
(P_W, f_1 t- H_1, \ldots, f_7 t - H_7).
\ee
We further impose the scaling relation associated to the new coordinate $t$
\be
\left(f_1,f_2,...,t\right) \sim \left(\lambda f_1, \lambda f_2 ,..., \lambda^{-1} t\right) \;.
\ee
We can then resolve the $SU(2)$ singularity over $T:\;t=0$ as before by $\left(x,y,t\right) \rightarrow \left(x r, y r, t r \right)$ and by imposing $\left(x,y,t,r\right) \sim \left(\lambda^{-1} x,\lambda^{-1} y,\lambda^{-1} t,\lambda r \right)$. The resulting space is now smooth over $T$ with an $I_2$ fibre over a generic point, while over certain points in $T$, denoted $\hat{C}_{II}$, the fibre will factorise to an $I_3$. The exceptional divisor $R:\;r=0$ forms the Cartan of the $SU(2)$ on the Coulomb branch. 

We can perform this blow-up and resolution in the deformed geometry $P_W$ which directly gives the final smooth space with torsion cycles. This simply amounts to dropping the restriction $e_4 = \frac{1}{4}b^2$.  However to identify the torsional 2-cycles using the arguments presented in the main text we need to work with the resolved geometry over $C_{II}$. Since the blow-up in the base is localised away from the locus $C_{I}$, it does not affect this locus. The crucial information is the intersection numbers of the sections with the fibre components over the points $\hat{C}_{II}$. These will allow us to identify the 3-chains that will, after the deformation, become the chains with a boundary of twice the torsional two-cycles. 

In principle this analysis can be done by using the computer package SINGULAR \cite{singular}, leading to a globally valid blowup and resolution of the singularities over $C_{II}$. However,
it is more instructive to perform a local analysis of the fibre over the $C_{II}$ which will be sufficient to extract the relevant intersection numbers with the sections. Our approach is to consider the locus given by $H_6=H_7=0$. This can be shown, by a prime decomposition, to be composed of the locus $C_{II}$ and the separate set of points $e_0=e_1=0$. We will ignore these points in our local analysis though they would lead to $SU(2)$ singularities over points in the base after the blow-up. Indeed since the set of points $C_{II}$ does not intersect the curve $e_0=0$ \cite{Morrison:2014era},
we can restrict our attention to the subset $e_0 \neq 0$, where in particular we can allow for functions meromorphic in $e_0$. We can now explicitly solve the two equations
\be
f_6 t - H_6 = 0 \;,\;\; f_7 t - H_7 = 0 \;,
\ee
which gives
\bea
e_3 &=& \frac{-e_1^3 + 4 e_0 e_1 e_2 + f_7 t}{8 e_0^2} , \\
b^2 &=& \frac{e_1^4 - 8 e_0 e_1^2 e_2 + 16 e_0^2 e_2^2 + 4 e_0 f_6 t - e_1 f_7 t}{16 e_0^3} \;. \nn
\eea
Since only $b^2$ appears in $P_W$ we can plug this back into the equation to analyse the fibre structure explicitly. This solution is valid away from $e_0=0$ and also away from $b=0$, where the coordinate change $\left(e_3,b\right) \rightarrow \left(f_6,f_7\right)$ degenerates. We now redefine
\be
x \rightarrow x + \frac{\left(-3 e_1^2 + 8 e_0 e_2\right) z^2}{12 e_0} \;
\ee
to bring the $SU(2)$ singularity over $T$ to $x=y=0$. Finally we resolve it by introducing $R:\;r=0$ as $\left(x,y,t\right) \rightarrow \left(x r, y r, t r \right)$. There are then two fibre components over the exceptional divisor in the base,
\bea
A_{II}\;&:&\;T \cap \hat{P}_{W} \cap \left\{C_{\mathrm{base}}\right\} \;, \nn \\
B_{II}\;&:&\;R \cap \hat{P}_{W} \cap \left\{C_{\mathrm{base}}\right\} \;. \label{su2comprsec}
\eea
The interesting $I_3$ locus can be identified from the discriminant to lie on
\bea \label{eq-hatCII}
\hat{C}_{II} \;:\; && \left\{  - 32 e_2  f_7^2 -16 e_0^2 f_6^2 + 24 e_0 e_1 f_6 f_7 + 3 e_1^2 f_7^2 = 0 \right\} \nn \\
  && \, \,  \cap \, \{t=0\}\;
\eea
(viewed as a locus on the base),
and over this locus the fibre component $B_{II}$ splits into components
\bea
B_{II,1} &:& \{8 f_7 y - 8 f_6 x z - 6 e_1 f_7 x z - f_7^2 t z^3 = 0\} \cap R \cap P_{\hat{C}_{II}}   \;, \nn \\
B_{II,2} &:& \{8 f_7 y + 8 f_6 x z + 6 e_1 f_7 x z + f_7^2 t z^3 = 0\} \cap R \cap  P_{\hat{C}_{II}}  \;\nn \nonumber 
\eea
with  $P_{\hat{C}_{II}}$ the divisor associated to the first polynomial in (\ref{eq-hatCII}).
Note that we have set $e_0=-1$ in the above for simplicity, and have given only the important component of the intersecting equations defining the fibre. The other component of the fibre over these points is 
\bea
A_{II} :  && \{16 r f_2^2 x^3 - 16 f_2^2 y^2 + 16 f_1^2 x^2 z^2 + 24 e_1 f_6 f_7 x^2 z^2  \nn \\ 
&& +  9 e_1^2 f_7^2 x^2 z^2  = 0\}    \cap T \cap P_{\hat{C}_{II}}    \;.
\eea
We can now intersect these components with the proper transform of the sections $U:\;z=0$ and $S:\;\left[x,y,z\right]=\left[e_3^2 - \frac{2}{3} b^2 e_2,- e_3^3 + b^2 e_2 e_3 - \frac{1}{2} b^4 e_1,i b \right ]$ \cite{Morrison:2012ei}, given here on the Weierstra\ss{} model before blowup and resolution,  which after some calculation eventually yields the intersection numbers 
\bea
U \cdot A_{II} &=& 1 \;,\;\; U \cdot B_{II,1} = 0 \;,\;\; U \cdot B_{II,2} = 0 \;,\nn\\
S \cdot A_{II} &=& 0 \;,\;\; S \cdot B_{II,1} = 1 \;,\;\; S \cdot B_{II,2} = 0 \;.  \\
R \cdot A_{II} &=& 2 \;,\;\; R \cdot B_{II,1} = -1 \;,\;\; R \cdot B_{II,2} = -1 \;.\nn
\eea
This identifies the component of the fibre which becomes the torsional 2-cycle after the deformation as $B_{II,1} - B_{II,2}$.

\bibliography{papers}  
\bibliographystyle{utphys}

\end{document}